\documentclass[%
 aip,
 apl,
 amsmath,amssymb,
 reprint,%
]{revtex4-1}
\usepackage{graphicx}
\usepackage{dcolumn}
\usepackage{bm}

\usepackage[utf8]{inputenc}
\usepackage[T1]{fontenc}
\usepackage{mathptmx}
\usepackage{etoolbox}

\usepackage{float}
\usepackage{xcolor}

\makeatletter
\def\@email#1#2{%
 \endgroup
 \patchcmd{\titleblock@produce}
  {\frontmatter@RRAPformat}
  {\frontmatter@RRAPformat{\produce@RRAP{*#1\href{mailto:#2}{#2}}}\frontmatter@RRAPformat}
  {}{}
}%
\makeatother
\begin{document}

\preprint{AIP/123-QED}

\title{Fourier\mbox{-}domain Crosstalk and Leakage Correction for Binary\mbox{-}Encoded Fourier Single\mbox{-}Pixel Imaging}
\author{Chongwu Shao}
\affiliation{
Center for Quantum Technology Research and Key Laboratory of Advanced Optoelectronic Quantum Architecture and Measurements, School of Physics, 
Beijing Institute of Technology, Beijing 100081, China
}

\author{Yue Cao}
\affiliation{
Chinese Academy of Agricultural Mechanization Sciences Group Co.,Ltd., 
Beijing 100083, China
}

\author{Wei Zhang}
\affiliation{
Center for Quantum Technology Research and Key Laboratory of Advanced Optoelectronic Quantum Architecture and Measurements, School of Physics, 
Beijing Institute of Technology, Beijing 100081, China
}

\author{Xiao-Peng Jin}
\affiliation{
Center for Quantum Technology Research and Key Laboratory of Advanced Optoelectronic Quantum Architecture and Measurements, School of Physics, 
Beijing Institute of Technology, Beijing 100081, China
}

\author{Yinran Shen}
\affiliation{
Center for Quantum Technology Research and Key Laboratory of Advanced Optoelectronic Quantum Architecture and Measurements, School of Physics, 
Beijing Institute of Technology, Beijing 100081, China
}

\author{Shijian Li*}
\affiliation{
School of Photonics and Optical Engineering, Aerospace Information Technology University,
Jinan 250200, China
}
\affiliation{
Shandong Key Laboratory of Intelligent Photonic Transmission and Sensing, Jinan 250200, China
}
\author{Xu-Ri Yao*}
\affiliation{
Center for Quantum Technology Research and Key Laboratory of Advanced Optoelectronic Quantum Architecture and Measurements, School of Physics, 
Beijing Institute of Technology, Beijing 100081, China
}
\affiliation{
State Key Laboratory of Chips and Systems for Advanced Light Field Display, School of Physics,
Beijing Institute of Technology, Beijing 100081, China
}



\begin{abstract}
Digital micromirror devices enable high-speed Fourier single\mbox{-}pixel imaging (Fourier\mbox{-}SPI), but their binary modulation introduces quantization errors that degrade the reconstructed image quality.
In particular, each binarized Fourier probe contains non\mbox{-}target Fourier components, thereby distorting the corresponding Fourier\mbox{-}domain measurement.
This Letter presents a Fourier\mbox{-}domain correction framework that characterizes these errors using the Fourier\mbox{-}domain response matrix of the decoded binary Fourier patterns.
Accordingly, a response\mbox{-}matrix\mbox{-}based Fourier\mbox{-}SPI measurement model is established by partitioning the matrix into components associated with the sampled and unsampled frequencies, thereby decomposing the quantization error into sampled\mbox{-}frequency crosstalk and unsampled\mbox{-}frequency leakage.
This model established a generalized correction framework, termed Fourier\mbox{-}domain Crosstalk and Leakage Correction (FCLC), alongside two simplified methods: Fourier\mbox{-}domain Sampled\mbox{-}Frequency Crosstalk Correction (FSCC) and Fourier\mbox{-}domain Direct\mbox{-}Current Crosstalk Correction (FDCC).
FSCC is implemented via an inversion\mbox{-}free Richardson\mbox{-}type iteration, whereas FDCC admits a closed\mbox{-}form solution.
Numerical simulations and experiments across multiple binary encoding strategies reveal that the proposed framework improves reconstruction fidelity without modifying the binary patterns or acquisition process.
\end{abstract}

\maketitle

Single\mbox{-}pixel imaging (SPI)~\cite{edgar2019principles,gibson2020single,song2025advances} reconstructs images from sequentially modulated illumination or detection patterns using a single\mbox{-}pixel detector. Among its variants, Fourier\mbox{-}SPI~\cite{zhang2015single,bian2016efficient,zhang2016three,tang2022fast} acquires Fourier coefficients through phase\mbox{-}shifted measurements and reconstructs images via the inverse Fourier transform (IFT). In this work, the reconstructed image is formulated as
\begin{equation}
\boldsymbol{x}
=
\mathcal{N}
\left\{
\left|
\operatorname{Re}
\left[
\operatorname{IFT}(\boldsymbol{f})
\right]
\right|
\right\},
\end{equation}
where \(\boldsymbol{f}\) denotes the Fourier coefficients used for reconstruction, \(\boldsymbol{x}\) denotes the normalized reconstructed image, and \(\mathcal{N}(\cdot)\) is the min--max normalization operator defined as
\(\mathcal{N}(\boldsymbol{u})=[\boldsymbol{u}-\min(\boldsymbol{u})]/[\max(\boldsymbol{u})-\min(\boldsymbol{u})]\).

\begin{figure}[H]
\centering
\includegraphics[width=1\columnwidth]
{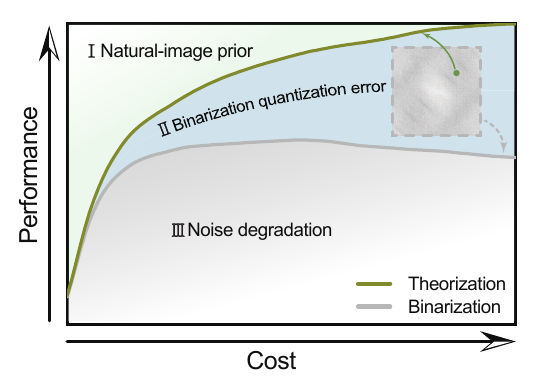}
\caption{
Illustration of the performance\mbox{-}cost trade\mbox{-}off in Fourier\mbox{-}SPI.
The two reference curves compare the IFT reconstruction fidelity of theorized and binarized Fourier\mbox{-}SPI.
The three shaded regions indicate potential fidelity gains from natural\mbox{-}image priors, performance degradation induced by binarization quantization errors, and further degradation caused by measurement noise.
The inset depicts binarization\mbox{-}induced spectral leakage, with the target\mbox{-}frequency response highlighted in green and the non\mbox{-}target frequency responses shown in gray.
}
\label{fig:background}
\end{figure}

Digital micromirror devices (DMDs) are widely employed for high\mbox{-}speed pattern modulation; however, their binary operation requires continuous\mbox{-}valued Fourier basis patterns to be approximated using a finite number of binary patterns. Regardless of the encoding strategy, this approximation inevitably introduces binarization\mbox{-}induced errors under practical encoding constraints. 
Specifically, the decoded probe deviates from its target Fourier basis and can be expressed as a weighted superposition of target and non\mbox{-}target Fourier components. 
During Fourier\mbox{-}SPI measurements, these non\mbox{-}target weights cause other Fourier coefficients to contaminate the measured target coefficient, thereby degrading reconstruction quality.
,as shown in Fig.~\ref{fig:background}.

To mitigate binarization\mbox{-}induced quantization errors and drive the performance\mbox{-}cost trade\mbox{-}off of Fourier\mbox{-}SPI closer to its idealized counterpart, two principal approaches are viable.
The first approaches focuses on designing advanced binary encoding strategies that suppress the magnitude and optimize the distribution of non\mbox{-}target frequency responses.
Representative strategies include increasing the number of binary patterns~\cite{huang2018computational}, optimizing error\mbox{-}diffusion kernels and scanning paths~\cite{liang2019fast,li2022full}, employing positive\mbox{-}negative dithering~\cite{fan2025enhanced}, adopting frequency\mbox{-}dependent diffusion optimization~\cite{ma2024physics,jiang2025globally}, and jointly generating \(K\) binary patterns via complex\mbox{-}domain error diffusion~\cite{shao2026geometry}.
The second approach addresses binarization\mbox{-}induced errors through post\mbox{-}processing.
In the spatial\mbox{-}domain, known binary patterns are exploited to estimate and correct the discrepancies between actual and ideal Fourier measurements during reconstruction~\cite{jiang2026binary}.
Alternatively, in the frequency\mbox{-}domain, the non\mbox{-}target frequency responses identified above can be explicitly modeled to enable direct correction of binarization\mbox{-}induced errors.

In this Letter, a measurement model for Fourier\mbox{-}SPI is established based on the non\mbox{-}target frequency responses induced by binary quantization. 
Specifically, each decoded Fourier measurement is formulated as the sum of the desired target\mbox{-}frequency response, crosstalk from other sampled Fourier coefficients, and leakage from unsampled spectral components.
Based on this model, the Fourier\mbox{-}domain Crosstalk and Leakage Correction (FCLC) framework is proposed to improve Fourier coefficient recovery by applying a correction operator directly to the decoded Fourier coefficients. 
Considering the low\mbox{-}frequency\mbox{-}prioritized sampling scheme in Fourier\mbox{-}SPI, Fourier\mbox{-}domain Sampled\mbox{-}Frequency Crosstalk Correction (FSCC) is introduced as a practical special case of FCLC, reducing computational cost while retaining much of its correction capability. A matrix\mbox{-}free Richardson\mbox{-}type iterative solver is designed for rapid convergence to the corrected Fourier coefficients. Further simplification yields Fourier\mbox{-}domain Direct\mbox{-}Current Crosstalk Correction (FDCC), providing a closed\mbox{-}form standalone correction and an initialization for the FSCC solver to meet the speed requirements of Fourier\mbox{-}SPI.
Numerical simulations across representative binary encoding strategies and Fourier\mbox{-}SPI experiments demonstrate that the proposed framework effectively mitigates binarization\mbox{-}induced quantization errors, improving Fourier coefficient recovery and reconstruction quality through Fourier domain postprocessing alone, without modifying the binary encoding or measurement process.

Following binarization and decoding, the probe corresponding to the \(i\)\mbox{-}th measurement, \(\tilde{\phi}_{i}\), inevitably deviates from its target Fourier basis and can be expanded in terms of the complete set of normalized Fourier bases \(\{\phi_j\}_{j=0}^{N-1}\) as
\begin{equation}
\tilde{\phi}_{i}
=
\sum_{j=0}^{N-1}
a_{ij}\phi_{j},
\label{eq:probe}
\end{equation}
where \(N\) denotes the total number of Fourier basis components and is equal to the number of image pixels, \(\phi_j\) denotes the \(j\)\mbox{-}th normalized Fourier basis, and \(a_{ij}\) is the corresponding complex coefficient of \(\phi_j\) in the decoded probe.

In Fourier\mbox{-}SPI, each single\mbox{-}pixel measurement is obtained by spatially integrating the product of the decoded probe and the target scene. By expanding both the decoded probe and the scene in terms of the normalized Fourier basis, The Fourier coefficient \(\tilde{f}_i\) obtained by the \(i\)\mbox{-}th Fourier probe can be expressed as
\begin{equation}
\begin{aligned}
\tilde{f}_i
=
\sum_{x=0}^{N-1}
\tilde{\phi}_{i}(x)I(x)
&=
\sum_{j=0}^{N-1}
\sum_{j'=0}^{N-1}
a_{ij}f_{j'}
\sum_{x=0}^{N-1}
\phi_j(x)\phi_{j'}(x)
\\
&=
\sum_{j=0}^{N-1}
a_{ij}
f_{(-j)\bmod N},
\end{aligned}
\label{eq:single_measurement}
\end{equation}
where \(I(x)\) denotes the target scene and \(f_j\) is its \(j\)\mbox{-}th Fourier coefficient. 
The last equality follows from the orthogonality of the Fourier basis, under which only opposite\mbox{-}frequency pairs survive the spatial summation.
Consequently, each probe measurement in binary\mbox{-}encoded Fourier\mbox{-}SPI can be formulated as a weighted linear combination of all Fourier coefficients of the target scene.

Under undersampling, only \(m\) target Fourier frequencies are sampled, yielding \(m\) equations of the form of Eq.~\eqref{eq:single_measurement}, which can be expressed in matrix form. 
By reordering the Fourier coefficients according to the opposite\mbox{-}frequency correspondence in the above equation, the Fourier\mbox{-}SPI process can be compactly written as
\begin{equation}
\tilde{\mathbf{f}}_{\mathrm{s}}
=
\mathbf{A}\mathbf{f},
\end{equation}
where \(\mathbf{f}\in\mathbb{C}^{N}\) denotes the vectorized Fourier spectrum arranged in the corresponding opposite\mbox{-}frequency order, and \(\tilde{\mathbf{f}}_{\mathrm{s}}\in\mathbb{C}^{m}\) denotes the Fourier coefficient vector directly decoded from the measurements. 
\(\mathbf{A}=[a_{ij}]\in\mathbb{C}^{m\times N}\) denotes the Fourier\mbox{-}domain response matrix of the binary probes, which becomes \(\mathbf{A}_{N}\in\mathbb{C}^{N\times N}\) under full\mbox{-}sampling conditions. 
The resulting linear system is underdetermined when \(m<N\) and square when \(m=N\).

To distinguish the contributions from sampled and unsampled frequencies, the Fourier coefficients are reordered by placing the \(m\) sampled\mbox{-}frequencies precede the remaining \(N-m\) unsampled\mbox{-}frequencies.
The spectrum vector and the Fourier\mbox{-}domain response matrix can then be partitioned as
\begin{equation}
\mathbf{f}
=
\begin{bmatrix}
\mathbf{f}_{\mathrm{s}}\\
\mathbf{f}_{\mathrm{u}}
\end{bmatrix},
\qquad
\mathbf{A}
=
\begin{bmatrix}
\mathbf{a} & \mathbf{b}
\end{bmatrix},
\end{equation}
where \(\mathbf{f}_{\mathrm{s}}\in\mathbb{C}^{m}\) contains the actively sampled Fourier coefficients, whereas \(\mathbf{f}_{\mathrm{u}}\in\mathbb{C}^{N-m}\) contains the remaining unsampled coefficients. Correspondingly, \(\mathbf{a}\in\mathbb{C}^{m\times m}\) denotes the sampled\mbox{-}frequency crosstalk matrix, which characterizes the coupling among the sampled Fourier coefficients, while \(\mathbf{b}\in\mathbb{C}^{m\times(N-m)}\) denotes the unsampled\mbox{-}frequency leakage matrix, which characterizes the contribution of the unsampled spectral components to the measurements.
The measurement model can therefore be rewritten as
\begin{equation}
\tilde{\mathbf{f}}_{\mathrm{s}}
=
\mathbf{a}\mathbf{f}_{\mathrm{s}}
+
\mathbf{b}\mathbf{f}_{\mathrm{u}},
\label{eq:model}
\end{equation}
which explicitly separates the sampled\mbox{-}frequency crosstalk from the unsampled\mbox{-}frequency leakage in binary\mbox{-}encoded Fourier\mbox{-}SPI.
Under ideal measurement conditions, the crosstalk matrix should satisfy \(\mathbf{a}=\mathbf{I}\), while the leakage matrix should satisfy \(\mathbf{b}=\mathbf{0}\). 
For a well\mbox{-}designed Fourier\mbox{-}SPI binary encoding strategy, the sampled\mbox{-}frequency response matrix \(\mathbf{a}\) is expected to be of full rank, ensuring the unique recovery of the sampled Fourier coefficients.

To recover the actively sampled Fourier coefficients from the measurements, a linear correction operator \(\mathbf{O}\in\mathbb{C}^{m\times m}\) is introduced such that \(\mathbf{O}\tilde{\mathbf{f}}_{\mathrm{s}}=\mathbf{f}_{\mathrm{s}}\).
Accordingly, the ideal operator \(\mathbf{O}\) should satisfy
\begin{equation}
\mathbf{O}\mathbf{a}
=
\mathbf{I},
\qquad
\mathbf{O}\mathbf{b}
=
\mathbf{0}.
\label{eq:ideal}
\end{equation}
However, these two ideal conditions generally cannot be satisfied simultaneously.
The ideal constraints are therefore relaxed by introducing a trade\mbox{-}off parameter \(\lambda\geq0\) to balance sampled\mbox{-}frequency crosstalk correction and unsampled\mbox{-}frequency leakage suppression. 
The practical correction operator \(\hat{\mathbf{O}}\) is then obtained from
\begin{equation}
\hat{\mathbf{O}}
=
\operatorname*{arg\,min}_{\mathbf{O}}
\left\{
\left\|
\mathbf{O}\mathbf{a}
-
\mathbf{I}
\right\|_{F}^{2}
+
\lambda
\left\|
\mathbf{O}\mathbf{b}
\right\|_{F}^{2}
\right\}.
\label{eq:FCLC_opt}
\end{equation}
Since \(\mathbf{a}\) is full rank, \(\mathbf{a}\mathbf{a}^{H}+\lambda\mathbf{b}\mathbf{b}^{H}\) is Hermitian positive definite for \(\lambda\geq0\), ensuring that the objective function is strictly convex and has a unique global minimizer. Setting its first\mbox{-}order optimality condition with respect to \(\mathbf{O}\) to zero gives \(2(\mathbf{O}\mathbf{a}-\mathbf{I})\mathbf{a}^{H}+2\lambda\mathbf{O}\mathbf{b}\mathbf{b}^{H}=0\). Since \(\mathbf{a}\mathbf{a}^{H}+\lambda\mathbf{b}\mathbf{b}^{H}\) is invertible, the unique solution is given by
\begin{equation}
\hat{\mathbf{O}}
=
\mathbf{a}^{H}
\left(
\mathbf{a}\mathbf{a}^{H}
+
\lambda\mathbf{b}\mathbf{b}^{H}
\right)^{-1}.
\label{eq:FCLC}
\end{equation}
The operator \(\hat{\mathbf{O}}\) provides the complete correction formulation within the FCLC framework.

In the special case of \(\lambda=1\), the correction operator \(\hat{\mathbf{O}}_{\lambda=1}\) simplifies to \(\mathbf{a}^{H}(\mathbf{A}\mathbf{A}^{H})^{-1}\).
For a full\mbox{-}row\mbox{-}rank measurement matrix \(\mathbf{A}\), the Moore--Penrose pseudoinverse \(\mathbf{A}^{\dagger}\) is given by \(\mathbf{A}^{H}(\mathbf{A}\mathbf{A}^{H})^{-1}\).
Accordingly, \(\hat{\mathbf{O}}_{\lambda=1}\) precisely coincides with the rows of \(\mathbf{A}^{\dagger}\) associated with the sampled Fourier coefficients.
Consequently, the recovered coefficients are the sampled components of the minimum\mbox{-}\(\ell_{2}\)\mbox{-}norm solution to the complete undersampled system~\cite{Penrose1955}, which fails to capture the structural characteristics of natural images.
This indicates excessive weighting of unsampled\mbox{-}frequency leakage suppression for image reconstruction, thereby motivating the choice of \(0\leq\lambda<1\).

\begin{figure}[H]
\centering
\includegraphics[width=1\columnwidth]
{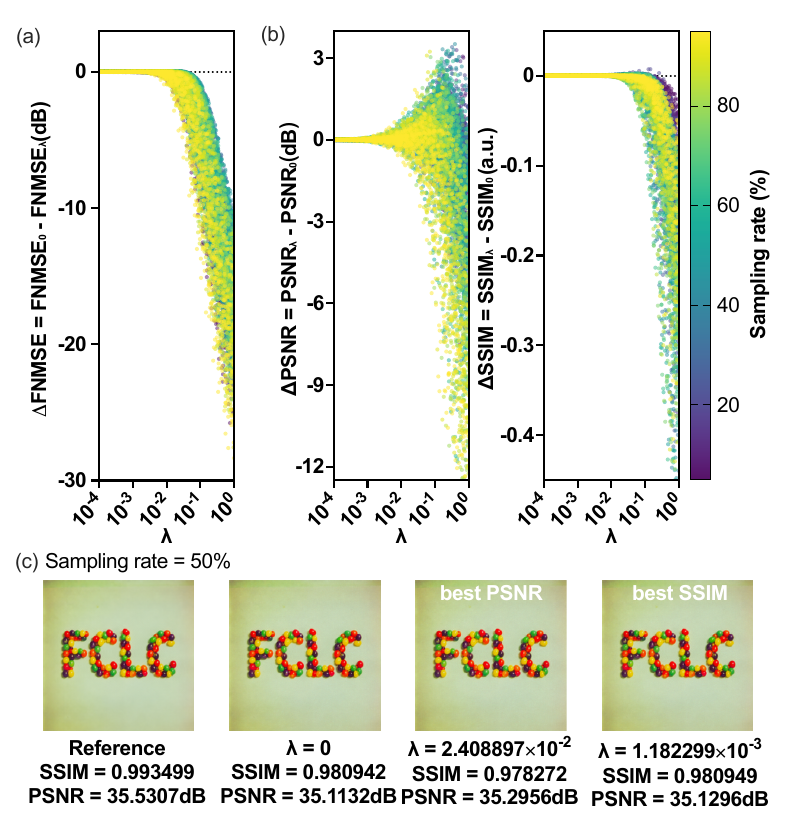}
\caption{Numerical simulation results illustrating the dependence of the FCLC framework on \(\lambda\). The correction operator \(\hat{\mathbf{O}}\) is computed using the conjugate\mbox{-}gradient (CG) method~\cite{Hestenes1952}.
(a) FNMSE gain relative to \(\lambda=0\) as a function of \(\lambda\), characterizing the improvement in Fourier\mbox{-}domain correction accuracy.
(b) PSNR and SSIM gains relative to \(\lambda=0\) as functions of \(\lambda\), characterizing the enhancement in image\mbox{-}domain reconstruction quality.
(c) Fourier\mbox{-}SPI reconstruction results for \(\lambda=0\), together with optimal results yielding the best PSNR and SSIM obtained by scanning \(\lambda\).
\\
\textit{Simulation settings:} Kodak24 dataset~\cite{kodak24}, \(256\times256\) pixels, four\mbox{-}step phase\mbox{-}shifting Floyd--Steinberg~\cite{floyd1976adaptive} binary encoding, without spatial upsampling.
}
\label{fig:lambda}
\end{figure}

The trade\mbox{-}off parameter \(\lambda\) admits a statistical interpretation from the spectral energy distribution of the target scene. For the low\mbox{-}frequency\mbox{-}prioritized sampling scheme considered here, the spectral energy is predominantly concentrated at low frequencies; consequently, an appropriate value of \(\lambda\) is expected to be substantially smaller than unity. 
The influence of \(\lambda\) on reconstruction performance was further investigated numerically, with \(\lambda=0\) serving as the baseline, as shown in Fig.~\ref{fig:lambda}.

The results in Fig.~\ref{fig:lambda}(a) indicate that tuning \(\lambda\) has a negligible effect on Fourier\mbox{-}domain recovery accuracy.
In Fig.~\ref{fig:lambda}(b), increasing \(\lambda\) leads to greater variability in PSNR, without a consistent improvement, while SSIM exhibits a marked degradation. 
Fig.~\ref{fig:lambda}(c) further illustrates the reconstructions obtained by individually optimizing \(\lambda\) separately for the maximum PSNR and SSIM, for which the improvement in reconstruction quality remains marginal.
These results indicate that, in low\mbox{-}frequency\mbox{-}prioritized Fourier\mbox{-}SPI, explicitly accounting for leakage from unsampled frequencies provides only limited practical benefit.

Given the marginal performance gain from explicitly accounting for unsampled\mbox{-}frequency leakage and the potentially adverse conditioning of the system matrix, doing so incurs unnecessary computational overhead. 
A practical approximation is therefore to set \(\lambda=0\), focusing the correction on the deterministic crosstalk among the actively sampled Fourier coefficients.
Accordingly, the correction operator \(\hat{\mathbf{O}}\) reduces to its simplified form
\begin{equation}
\hat{\mathbf{O}}^{\prime}
=
\mathbf{a}^{-1}.
\label{eq:FSCC}
\end{equation}
The correction operator \(\hat{\mathbf{O}}^{\prime}\) defines FSCC, a practically useful special case of the FCLC framework.

The FSCC can therefore be directly formulated as the linear system
\begin{equation}
\mathbf{a}\hat{\mathbf{f}}_{\mathrm{s}}
=
\tilde{\mathbf{f}}_{\mathrm{s}}.
\end{equation}
To avoid explicit matrix inversion, the correction is implemented using a matrix\mbox{-}free iterative Richardson\mbox{-}type solver~\cite{saad2003iterative}. The sampled Fourier coefficients are iteratively refined as
\begin{equation}
{\hat{\mathbf{f}}_{\mathrm{s}}}^{(t+1)}
=
{\hat{\mathbf{f}}_{\mathrm{s}}}^{(t)}
+
\mathbf{W}_{\mathbf{a}}
\left(
\tilde{\mathbf{f}}_{\mathrm{s}}
-
\mathbf{a}{\hat{\mathbf{f}}_{\mathrm{s}}}^{(t)}
\right),
\label{eq:simplified_iteration}
\end{equation}
where the optimized diagonal weighting matrix \(\mathbf{W}_{\mathbf{a}}\) is defined as
\begin{equation}
\mathbf{W}_{\mathbf{a}}
=
\operatorname{diag}
\left(
\frac{a_{11}^{*}}{\sum_{j}|a_{1j}|^{2}},
\frac{a_{22}^{*}}{\sum_{j}|a_{2j}|^{2}},
\ldots,
\frac{a_{mm}^{*}}{\sum_{j}|a_{mj}|^{2}}
\right).
\end{equation}
To obtain a computationally efficient initial estimate for the iterative solver, the sampled\mbox{-}frequency crosstalk matrix \(\mathbf{a}\) is approximated by retaining only its diagonal response and dominant direct\mbox{-}current (DC) coupling.
Accordingly, the simplified correction operator \(\hat{\mathbf{O}}^{\prime}\) is further approximated, yielding
\begin{equation}
\hat{\mathbf{O}}^{\prime\prime}
=
\left(
\mathbf{D}_{\mathbf{a}}
+
\mathbf{a}_{\mathrm{DC}}\mathbf{e}_{1}^{T}
\right)^{-1}.
\label{eq:FDCC}
\end{equation}
where \(\mathbf{D}_{\mathbf{a}}=\operatorname{diag}(a_{11},\ldots,a_{mm})\) represents the diagonal response, \(\mathbf{a}_{\mathrm{DC}}=[0,a_{21},a_{31},\ldots,a_{m1}]^{T}\) describes the off\mbox{-}diagonal DC coupling and \(\mathbf{e}_{1}\) denotes the first standard basis vector corresponding to the DC component.
The correction operator \(\hat{\mathbf{O}}^{\prime\prime}\) defines FDCC, a further simplification of the FCLC framework that can also serve as a standalone correction method.
The resulting simplified system provides a closed\mbox{-}form estimate that serves as the initial condition for the subsequent iteration
\begin{equation}
\hat{\mathbf{f}}_{\mathrm{s}}^{(0)}
=
\mathbf{D}_{\mathbf{a}}^{-1}
\left(
\tilde{\mathbf{f}}_{\mathrm{s}}
-
\mathbf{a}_{\mathrm{DC}}f_{\mathrm{DC}}
\right),
\label{eq:O_FDCC}
\end{equation}
where \(f_{\mathrm{DC}}\) denotes the corresponding DC Fourier coefficient.

\begin{figure}[H]
\centering
\includegraphics[width=1\columnwidth]
{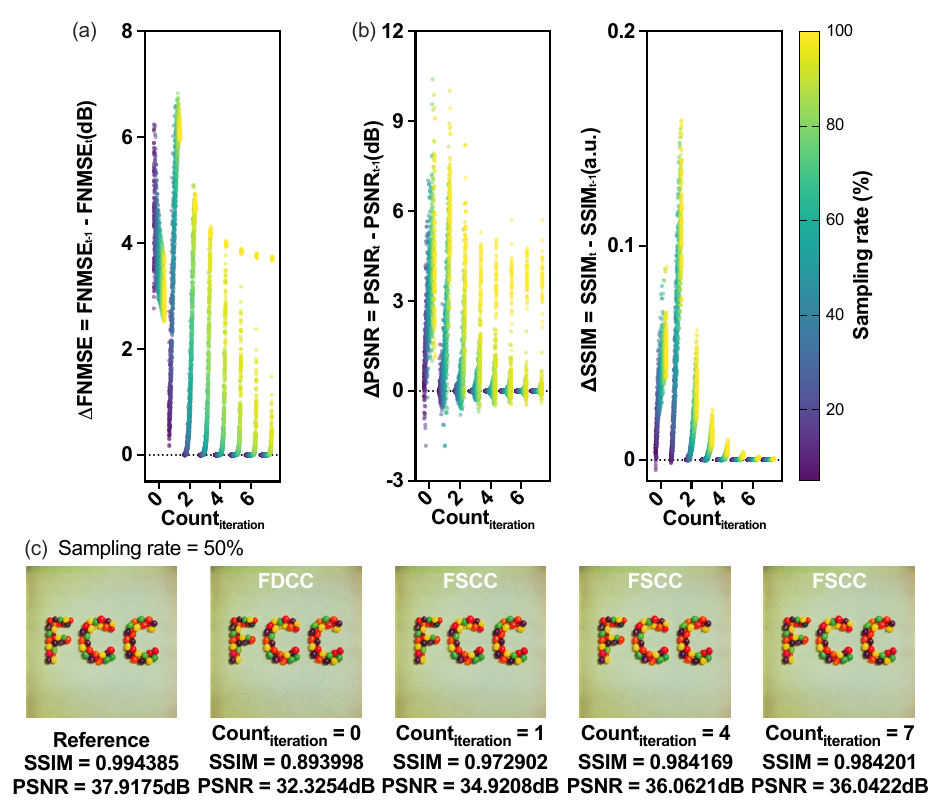}
\caption{Numerical simulation results of the Richardson\mbox{-}type iterative solver for FSCC, illustrating the incremental improvements in Fourier\mbox{-}spectrum correction accuracy and reconstruction performance across successive iterations.
(a) FNMSE gains achieved at each iteration, characterizing the progressive improvement in Fourier\mbox{-}spectrum correction accuracy.
(b) PSNR and SSIM gains achieved at each iteration, characterizing the progressive enhancement in image\mbox{-}domain reconstruction quality.
(c) Fourier\mbox{-}SPI reconstruction results obtained after 0, 1, 4, and 7 iterations, where the zero\mbox{-}iteration result corresponds to the FDCC initialization.
\\
\textit{Simulation settings:} Kodak24 dataset, \(256\times256\) pixels, four\mbox{-}step phase\mbox{-}shifting Floyd--Steinberg binary encoding, without spatial upsampling.
}
\label{fig:iteration}
\end{figure}

Defining the iteration error as
\(\boldsymbol{\varepsilon}^{(t)}
=
\hat{\mathbf{f}}_{\mathrm{s}}^{(t)}
-
\mathbf{f}_{\mathrm{s}}\),
its propagation is governed by
\begin{equation}
\boldsymbol{\varepsilon}^{(t+1)}
=
\mathbf{E}
\boldsymbol{\varepsilon}^{(t)},
\qquad
\mathbf{E}
=
\mathbf{I}
-
\mathbf{W}_{\mathbf{a}}\mathbf{a}.
\end{equation}
Accordingly, the iteration converges when
\(\rho(\mathbf{E})<1\) (Supplementary Note~1).
It is worth emphasizing that numerical convergence does not necessarily guarantee optimal reconstruction quality. Specifically, the neglected leakage from unsampled Fourier coefficients introduces model mismatch and degrades the converged solution.
The incremental improvement in coefficient recovery accuracy and image quality metrics between successive iterations were evaluated at different sampling rates, as shown in Fig.~\ref{fig:iteration}.

Figures~\ref{fig:iteration}(a) and (b) demonstrate that the initial few iterations consistently provide substantial improvements in both Fourier\mbox{-}domain accuracy and image quality, while Subsequent iterations yield only marginal gains.
Fig.~\ref{fig:iteration}(c) further presents representative reconstructions obtained at selected iteration numbers.
Therefore, a robust and conservative adaptive stopping criterion is essential to guarantee that the iterative process terminates once binarization\mbox{-}induced high\mbox{-}frequency artifacts have been sufficiently suppressed.

To this end, the iteration is terminated when the high\mbox{-}frequency TV of the IFT\mbox{-}reconstructed image \(\hat{\mathbf{x}}\) begins to increase or when the maximum number of iterations \(T_{\max}\) is reached. The stopping criterion is defined as
\begin{equation}
\begin{aligned}
\operatorname{TV}\!\left[
\mathcal{H}\!\left(\hat{\mathbf{x}}^{(T+1)}\right)
\right]
>
\operatorname{TV}\!\left[
\mathcal{H}\!\left(\hat{\mathbf{x}}^{(T)}\right)
\right]
\quad \lor \quad
T \geq T_{\max} = 7.
\end{aligned}
\label{eq:tv_stop}
\end{equation}
Where \(\mathcal{H}(\cdot)\) denotes the Gaussian high\mbox{-}pass filtering operator and \(\operatorname{TV}(\cdot)\) denotes the total variation~\cite{rudin1992nonlinear}. When the high\mbox{-}frequency TV begins to increase, the reconstruction at iteration \(T\) is retained as the final result.

\begin{figure}[H]
\centering
\includegraphics[width=1\columnwidth]
{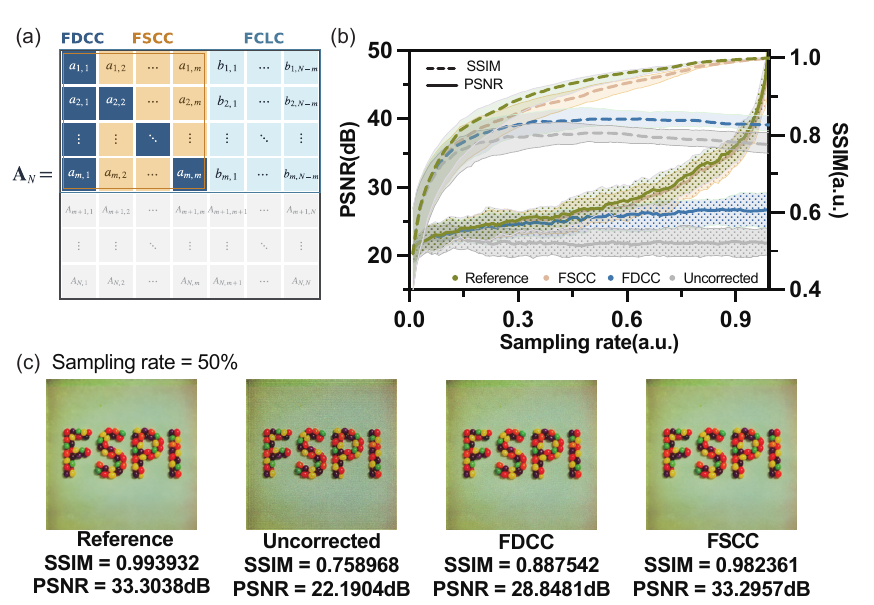}
\caption{Numerical simulation of IFT reconstruction correction in Fourier\mbox{-}SPI under the FCLC framework.
(a) Active components of the complete Fourier\mbox{-}domain response matrix \(\mathbf{A}\) for FCLC, FSCC, and FDCC.
(b) PSNR and SSIM comparison of IFT reconstructions based on FSCC, FDCC, and uncorrected coefficients on the Kodak24 dataset. Results are reported as median values, with \(Q_{25}\)--\(Q_{75}\) indicating the interquartile range (IQR). Reconstructions from the original Fourier basis patterns serve as the performance reference.
(c) Fourier\mbox{-}SPI reconstruction results of the Jelly Beans image obtained via IFT reconstruction based on FSCC, FDCC, and uncorrected coefficients.
\\
\textit{Simulation settings:} \(256\times256\) pixels, four\mbox{-}step phase\mbox{-}shifting Floyd--Steinberg binary encoding, without spatial upsampling.
}
\label{fig:perfermance_simulation}
\end{figure}

Consequently, FSCC and FDCC constitute two successive simplifications within the FCLC framework, progressively narrowing the correction scope and computational complexity.
To evaluate the practical effectiveness of the two simplified methods, FSCC and FDCC, their Fourier\mbox{-}SPI reconstruction performance was examined through numerical simulations, as shown in Fig.~\ref{fig:perfermance_simulation}.

Figure~\ref{fig:perfermance_simulation}(a) illustrates the progressive simplification of the FCLC framework from FSCC to FDCC.
The results in Fig.~\ref{fig:perfermance_simulation}(b) and (c) show that, for the basic Floyd--Steinberg binary encoding strategy, FDCC provides a substantial improvement in reconstruction quality, while FSCC further brings the performance close to the theoretical performance limit.

Notably, the proposed FCLC framework is agnostic to the choice of binarization strategy, provided that the sampled\mbox{-}frequency crosstalk matrix \(\mathbf{a}\) remains nonsingular.
In practice, FSCC employs a Richardson\mbox{-}type iterative solver for rapid convergence (Eq.~\ref{eq:simplified_iteration}). Although not unconditionally convergent, this solver can be substituted with a more robust alternative. Conversely, FDCC admits a closed\mbox{-}form solution (Eq.~\ref{eq:O_FDCC}), requiring only non\mbox{-}zero target frequency responses.
Typically, a higher\mbox{-}fidelity binary encoding yields a higher target\mbox{-}to\mbox{-}non\mbox{-}target response ratio, bringing the normalized full\mbox{-}sampling matrix \(\mathbf{A}_{N}\) closer to the identity matrix \(\mathbf{I}\), thereby favoring full rank, improving numerical conditioning, and potentially facilitating FSCC convergence.

To further evaluate the practical effectiveness of the proposed FCLC framework, real\mbox{-}object Fourier\mbox{-}SPI experiments were conducted using two color Zelda illustrations as imaging scene. 
Experiments were performed using a conventional passive DMD\mbox{-}based Fourier\mbox{-}SPI system with dichroic RGB separation and single\mbox{-}pixel detection (Supplementary Note~3).

Without ground truth, real\mbox{-}experiment reconstructions are evaluated using two customized no\mbox{-}reference metrics: the scene\mbox{-}specific natural image quality evaluator (SS\mbox{-}NIQE) and the wavelet\mbox{-}based median noise index (WMNI).
Adapted from the NIQE framework~\cite{mittal2012making}, SS\mbox{-}NIQE quantifies the statistical deviation between the reconstructed image and the original scene, whereas WMNI estimates local high\mbox{-}frequency noise in a smooth region using a one\mbox{-}level wavelet decomposition and median absolute deviation~\cite{donoho1995noising,chang2000adaptive} (see Supplementary Note~4).
Lower values of both metrics indicate better reconstruction quality.

\begin{figure}[H]
\centering
\includegraphics[width=1\columnwidth]{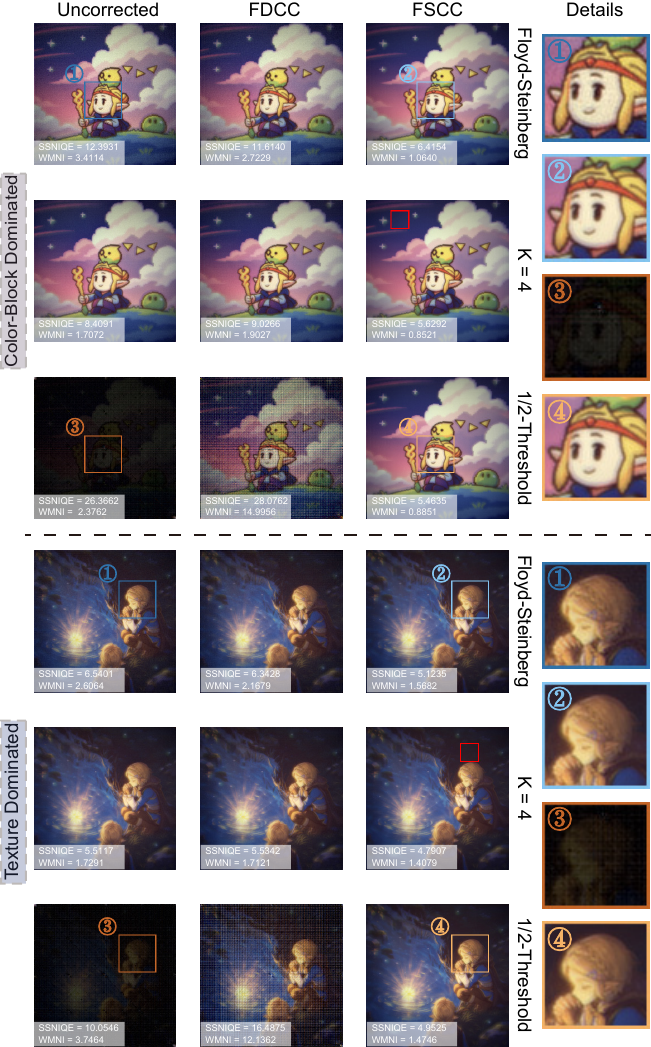}
\caption{Experimental validation of the proposed FCLC framework in real\mbox{-}object Fourier\mbox{-}SPI.
The three binary encoding strategies are four\mbox{-}step phase\mbox{-}shifting Floyd--Steinberg error diffusion, complex\mbox{-}domain error diffusion with \(K=4\)~\cite{shao2026geometry}, and four\mbox{-}step phase\mbox{-}shifting \(1/2\)\mbox{-}threshold binarization.
Reconstruction quality is evaluated using SS\mbox{-}NIQE and WMNI, with WMNI calculated over the same smooth region in all reconstructed images, as indicated by the red box in the \(K=4\) FSCC reconstruction.
\\
\textit{Experimental settings:} \(256\times256\) pixels, \(50\%\) sampling rate, four binary patterns for each complex Fourier basis pattern, without spatial upsampling.}
\label{fig:zelda}
\end{figure}

As illustrated in Fig.~\ref{fig:zelda}, the FCLC framework effectively corrects Fourier\mbox{-}domain measurement errors. Specifically, FDCC improves visual quality, while FSCC further enhances image clarity.
The improvement is particularly pronounced for \(1/2\)\mbox{-}threshold binarization, for which the uncorrected direct reconstruction is barely recognizable, whereas FSCC recovers a high\mbox{-}quality image.

For complex\mbox{-}domain error\mbox{-}diffusion encoding with \(K=4\), higher encoding fidelity enables direct reconstruction to outperform FDCC\mbox{-}corrected Floyd--Steinberg encoding, indicating that high\mbox{-}quality binary encoding remains important for applications with stringent real\mbox{-}time requirements.

In terms of computational overhead, FDCC provides closed\mbox{-}form correction with an additional processing time of only \(\sim4.74~\mathrm{ms}\), whereas each FSCC iteration requires \(\sim32.37~\mathrm{ms}\).
Higher\mbox{-}fidelity binary encoding generally leads to faster convergence, requiring fewer iterations to satisfy the stopping criterion.
The correction effectiveness of the FCLC framework across different sampling rates is further validated through experimental Fourier\mbox{-}SPI reconstructions (see Supplementary Note~5).

In summary, this Letter identifies binarization\mbox{-}induced non\mbox{-}target frequency responses as a source of reconstruction degradation in DMD\mbox{-}based Fourier\mbox{-}SPI, manifested as sampled\mbox{-}frequency crosstalk and unsampled\mbox{-}frequency leakage. Based on this mechanism, the FCLC framework is established for Fourier\mbox{-}domain error correction, with FSCC and FDCC providing two successive simplifications.
The normalized full\mbox{-}sampling Fourier\mbox{-}domain response error matrix \(\mathbf{D}_{\mathbf{A}_{N}}^{-1}\mathbf{A}_{N}-\mathbf{I}\) further provides a quantitative characterization of the inter\mbox{-}frequency coupling induced by various binary encoding strategies.
The full rank of \(\mathbf{a}\) guarantees a unique FCLC correction operator. For a rank\mbox{-}deficient \(\mathbf{a}\), a reduced full\mbox{-}rank response matrix can be constructed by selecting linearly independent measurement responses and appropriate Fourier components, enabling the application of the FCLC formulation to the reduced system.
Overall, the FCLC framework offers broad applicability across binary encoding strategies through Fourier Fourier domain postprocessing, without requiring modifications to the encoding or Fourier\mbox{-}SPI measurement process.

\hspace*{\fill}

See Supplementary Material for theoretical analysis and numerical validation of the general applicability of the FCLC framework, noise robustness analysis and noise\mbox{-}optimized correction operators, experimental configurations, full\mbox{-}reference and no\mbox{-}reference image quality metrics, and additional experimental results at different sampling rates.

\hspace*{\fill}

This work was supported by National Natural Science Foundation of China (Grant No.12504337) and Beijing Institute of Technology Research Fund Program for Young Scholars (202122012).

\section*{Author Declarations}
\subsection*{Conflict of Interest}
The authors declare no conflict of interest to disclose.

\subsection*{Author Contributions}
C.S. and S.L. conceived the research idea and developed the theoretical analysis.
C.S. and Y.C. performed the numerical simulations.
C.S., Y.C., X.Y., S.L., and W.Z. carried out the experiments.
C.S., Y.C., X.Y., S.L., W.Z., X.J., and Y.S. discussed the results.
C.S. and Y.C. prepared the manuscript, with revisions from all authors.
All authors reviewed and approved the final manuscript.

\section*{Data Availability Statement}

The Kodak24 dataset used in the numerical simulations is publicly available at \url{https://r0k.us/graphics/kodak/}.

The raw data and code supporting the findings of this study are available from the corresponding authors upon reasonable request.

\bibliographystyle{apsrev4-1}
\bibliography{references}

\end{document}